\documentclass[11pt,a4paper]{article} 
\pdfoutput=1
\usepackage{mathtools}
\usepackage[utf8x]{inputenc}     % if unicode errors are coming
\usepackage{jheppub}
\usepackage{enumerate}
\usepackage{epsfig} 
\usepackage{float} 
\usepackage[caption = false]{subfig}
\usepackage{wasysym} 
\usepackage{mathrsfs} 
\usepackage{amsfonts} 
\usepackage{amsbsy} 
\usepackage{amscd} 
\usepackage{pstricks} 
\usepackage{multirow} 
\usepackage{tikz}
\usepackage{color}
\usepackage{slashed}
\usepackage{array}
\usepackage{tablefootnote}

\usetikzlibrary{arrows,positioning,shapes.geometric} 
\usepackage[compat=1.1.0]{tikz-feynman}          % package allowing Feynman diagrams
\usepackage[font=small,labelfont=bf]{caption}
\usepackage{slashed}
\usepackage{multirow}
\usepackage{tabularx}
\usepackage{soul}  % Strikethrough text with \st{Hellow world}
\usepackage{listings} 
\usepackage{color}
\usepackage{amsthm}

\title{Hypergraphs in LHC Phenomenology$-$ The Next Frontier of IRC-Safe Feature Extraction}

\author[a]{Partha Konar,}
\author[a]{Vishal~S.~Ngairangbam,}
\author[b,c]{and Michael~Spannowsky} 

\affiliation[a]{Theoretical Physics Division, Physical Research Laboratory,\\ Shree Pannalal Patel Marg, Ahmedabad - 380009, Gujarat, India}
\affiliation[b]{Institute for Particle Physics Phenomenology, Durham University,\\ Durham DH1 3LE, United Kingdom}
\affiliation[c]{Department of Physics, Durham University,\\ Durham DH1 3LE, United Kingdom}
\emailAdd{konar@prl.res.in}
\emailAdd{vishalng@prl.res.in}
\emailAdd{michael.spannowsky@durham.ac.uk}

\abstract{In this study, we critically evaluate the approximation capabilities of existing infra-red and collinear (IRC) safe feature extraction algorithms, namely Energy Flow Networks (EFNs) and Energy-weighted Message Passing Networks (EMPNs). Our analysis reveals that these algorithms fall short in extracting features from any $N$-point correlation that isn't a power of two, based on the complete basis of IRC safe observables, specifically C-correlators. To address this limitation, we introduce the Hypergraph Energy-weighted Message Passing Networks (H-EMPNs), designed to capture any $N$-point correlation among particles efficiently. Using the case study of top vs. QCD jets, which holds significant information in its 3-point correlations, we demonstrate that H-EMPNs targeting up to N=3 correlations exhibit superior performance compared to EMPNs focusing on up to N=4 correlations within jet constituents. }

\preprint{IPPP/23/53}

\allowdisplaybreaks

\begin{document}
	\maketitle

\flushbottom
\section{Introduction} 

The Large Hadron Collider (LHC) has been a cornerstone in advancing our understanding of particle physics. However, the complexity of the data generated necessitates sophisticated methods for feature extraction and analysis. Traditional approaches often fail to capture intricate relationships among the data points, especially when considering infrared and collinear (IRC) safe observables. In this context, neural networks have shown promise~\cite{Andreassen:2019cjw,Komiske:2019fks,Bieringer:2020tnw,Kim:2021pcz,Lai:2021ckt, Romero:2021qlf,Batson:2021agz,Chahrour:2021eiv,Butter:2022xyj,Dreyer:2022yom,Onyisi:2022hdh,Kassabov:2023hbm,Butter:2023fov,Shen:2023ofd,Rousselot:2023pcj} but are not without limitations. These include issues regarding interpretibility~\cite{Andreassen:2018apy,Choi:2018dag,Dreyer:2020brq,Faucett:2020vbu,Lai:2020byl,Bogatskiy:2023nnw,Athanasakos:2023fhq}, uncertainty
quantification~\cite{Bollweg:2019skg,Ghosh:2021roe,Gambhir:2022gua,Butter:2021csz,dAgnolo:2021aun,Bellagente:2021yyh,Ghosh:2021hrh,Ghosh:2022lrf}, and a better handle and design of the physical biases~\cite{Komiske:2018cqr,Bogatskiy:2020tje,Kasieczka:2020nyd,Badger:2020uow,Maitre:2021uaa,Konar:2021zdg,Gong:2022lye,Hao:2022zns,Park:2022zov,Atkinson:2022uzb} of the neural networks for better physics generalization capabilities. The intricate nature of the underlying physical description warrants a thorough understanding of these algorithms, particularly as a precise understanding of the Standard Model background within perturbative Quantum Chromodynamics (pQCD) is needed to discover new physics.

\begin{figure}
	\centering
	\includegraphics[scale=0.25]{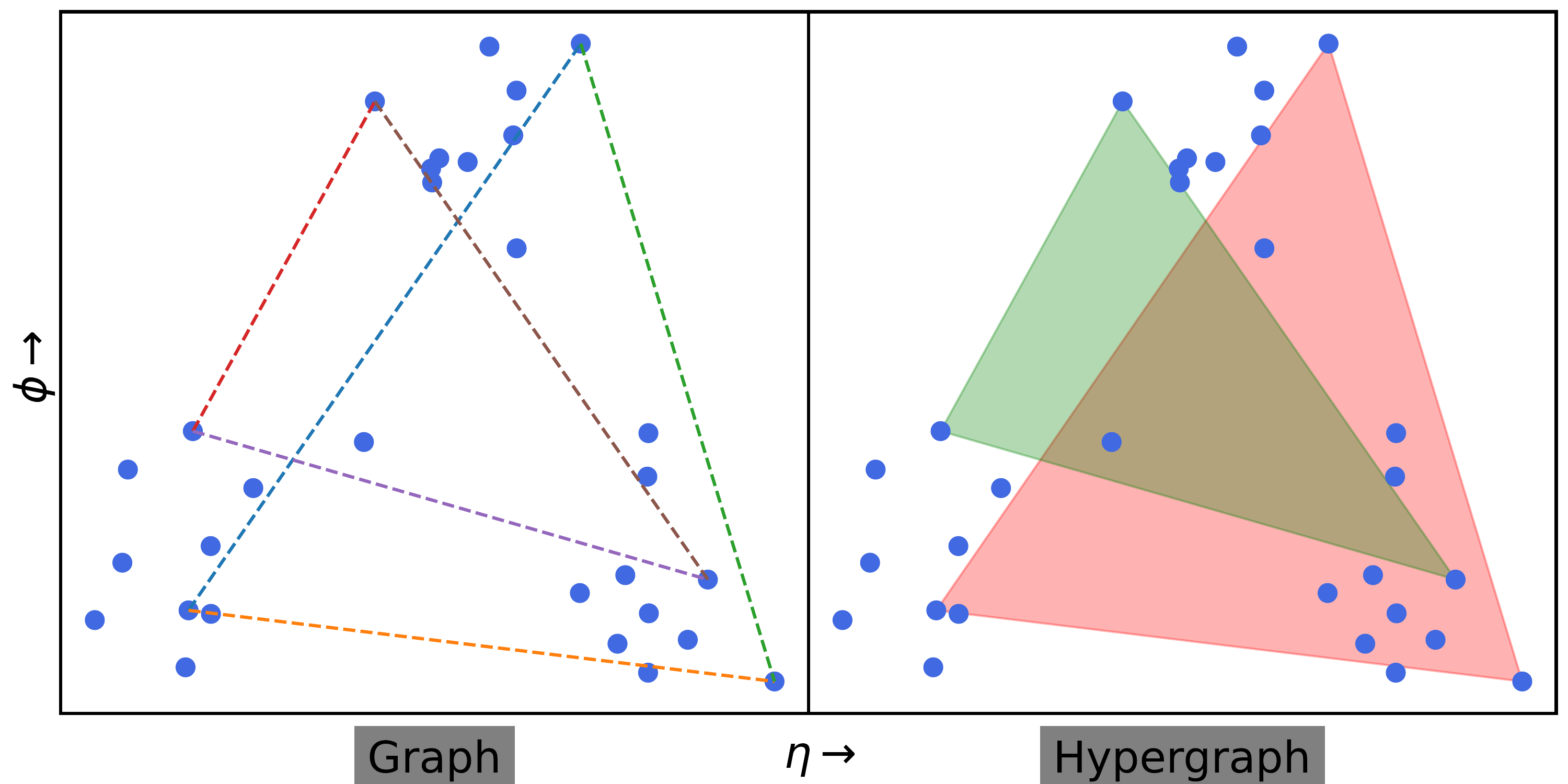} 
	\caption{Visualisation of the inter-relations of jet constituents as captured by a graph structure (left) and a hypergraph structure with order-three hyperedges (right). In a graph structure, the edges correlate two constituents at a time and are shown as a line segment connecting two nodes. Instead, the order-three hyperedges simultaneously link properties of three jet constituents at a time and are shown as a triangle with vertices coinciding with three nodes. Thus, hypergraphs are more expressive structures and can access higher-order correlations amongst jet constituents.}
	\label{fig:graph_vs_hgraph} 
\end{figure}

With the recorded events naturally represented as sets (of variable sizes) of different reconstructed particles or raw detector hits, point clouds are the natural representation of the recorded data, and architectures to process such data efficiently, particularly Graph Neural Networks~\cite{Qu:2019gqs,Mikuni:2020wpr,Bernreuther:2020vhm,Blance:2020ktp,Moreno:2019bmu,Atkinson:2021nlt,Atkinson:2021jnj,Tsan:2021brw}, have been used successfully for LHC phenomenology.
However, graphs do not expose higher-order correlations within the data by design, concentrating on two-particle correlations--the natural generalisation being hypergraphs. This generalisation is diagrammatically shown in figure~\ref{fig:graph_vs_hgraph} for a three-prong top jet where the graph's edges are defined in terms of two particles, while the order three hyperedges can look into the relevant three-prong structure of the top jet. This paper addresses these challenges by introducing Hypergraph Energy-weighted Message Passing Networks (H-EMPNs) designed to extract three-particle correlations better than existing IRC-safe feature extractors.  We first examine the universal approximation capabilities of existing infra-red and collinear safe neural network models like Energy Flow Networks (EFNs)~\cite{Komiske:2018cqr} and Energy-weighted Message Passing Networks (EMPNs)~\cite{Konar:2021zdg} in approximating any IRC safe observable expressible in terms of C-correlators~\cite{Tkachov:1995kk,Komiske:2017aww} looking into any general $N$-body phase space. Finding that EFNs are restricted $N=1$, and EMPNs have an arguably weak capability for approximating any $N\neq2^n$ C-correlators, we present H-EMPN as a more robust and versatile model capable of efficiently approximating any general IRC safe observable for any general $N$. Our method leverages the power of message-passing in graphs and hypergraphs to capture higher-order relationships among the data points, thereby providing a more comprehensive feature extraction mechanism.

Restricting ourselves to $N=3$ for the top vs QCD jet tagging scenario, where the dominant information lies in the 3-body decay phase space of the top quark, we find that H-EMPNs outperform EMPN, which look up to $N=4$ interparticle correlations, confirming our initial observation. We demonstrate the efficacy of H-EMPNs through empirical tests to showcase the learned graph representations. Furthermore, we discuss the architectural nuances of H-EMPN, providing insights into its design and training procedures. By doing so, we aim to establish H-EMPN as a powerful tool for LHC phenomenology, opening new avenues for applications in collider phenomenology.

Specifically, in section~\ref{sec:ua_irc}, we discuss the universal approximation of any IRC safe observable by EFNs and EMPNs by taking its correspondence to any generic C-correlator. In section.~\ref{sec:hepmn}, we devise H-EMPNs that can approximate any general C-correlator. The architecture and training details are presented in section~\ref{sec:arch}, while the results are presented in section.~\ref{sec:results}. We conclude in section~\ref{sec:conc}. 

\subsection*{Notation}
In the following discussions, we are given the set of four vectors of the jet constituents $$\mathcal{S}=\{\ p_1,p_2,....,p_{n_{part.}}\}\quad,$$ with $n_{part}$ being the number of constituents. These particles will be indexed via small Roman subscripts, while the number of message-passing operations will be indexed as Greek superscripts. Unless otherwise stated, all summations will be over the set $\mathcal{S}$. The four vectors are given in terms of the relative hardness $z_i=p_T^i/\sum p_T^j$ and the rapidity-azimuth variables $\mathbf{\hat{p}}_i=(y_i,\phi_i)$. 
Bold-faced alphabets like $\mathbf{h}_i$ and $\mathbf{G}$ denote vector quantities with their italicised counterparts $h_i$ and $G$ acting as a placeholder for a component. As we will consider inference on networks after training rather than the training itself, we will not explicitly write the dependence of function approximators on the tunable parameters. For instance,  $\mathbf{g}^{(\alpha)}(\mathbf{h}^{(\alpha-1)}_i,\mathbf{h}^{(\alpha-1)}_j)$ denotes a MultiLayer Perceptron (MLP) at the $\alpha^{th}$ message passing step, $\mathbf{h}^{(\alpha-1)}_i$ and $\mathbf{h}^{(\alpha-1)}_j$ correspond to the updated node features in the previous operation of particle $i$ and $j$, respectively, in $\mathcal{S}$.

\section{Universal Approximation of IRC safe observables} 
\label{sec:ua_irc} 
In present scientific literature, it is well-known that MLPs are universal function approximators~\cite{Cybenko1989,LESHNO1993861,arora2018understanding}. Without going into mathematical rigour, a parametrized function $f(\mathbf{x},\Theta)$ of a vector $\mathbf{x}$ and tunable parameters $\Theta$, is a universal approximator if it can approximate any continuous function up to any arbitrary precision in a compact domain and range. On the other hand, physical observables like momenta or position live in an underlying metric space, and notions of completeness have long been the bread-and-butter of physicists to study physical systems. The complete set of IRC safe observables is essential at the LHC and the subject of our present investigation. Any IRC safe observable $\mathcal{O}$ can be expanded in a basis of C-correlators~\cite{Tkachov:1995kk} as
\begin{equation}
	\label{eq:c_corr} 
	\mathcal{O}\approx \sum_{N=0}^{N_{max}}\; \mathcal{C}^{f_N}_N\quad,\quad \mathcal{C}^{f_N}_N=\sum_{i_1}\sum_{i_2}...\sum_{i_N}\, E_{i_1}E_{i_2}...E_{i_N}\,f_N(\hat{p}_{i_1},\hat{p}_{i_2},...\hat{p}_{i_N})\quad,
\end{equation}where $f_N$ is symmetric to any permutation of its arguments. 
Energy Flow Polynomials (EFPs)~\cite{Komiske:2017aww} expand $\mathcal{O}$ in a basis of polynomials of energy using the Stone-Weierstrass approximation theorem. In this section, we take a look into the approximation capabilities of existing IRC safe neural networks, namely Energy Flow Networks~\cite{Komiske:2018cqr}, and Energy-weighted Message Passing Network (EMPN)~\cite{Konar:2021zdg}, comparing the functional form to any arbitrary $N$ in the basis of C-correlators.  As the C-correlators are complete, the network-extracted observables would be expressible as a linear sum of different C-correlators, and we investigate the terms in the sum (as given in eq.~\ref{eq:c_corr}) that are optimally extracted via these observables.

Although we rely on the statement of universal approximation theorems, it is important to remember that we will strictly talk about the existence of such approximators and not concentrate on the method of finding such a function. However, presently available gradient descent algorithms are powerful enough to efficiently find an approximation given that we have the desired output value on a large enough number of samples. This numerical nature of finding a practical working point in the weight space is one of the significant concerns regarding the interpretability of neural networks in general. Our aim is not to tackle this more difficult problem but to systematically establish the capability of IRC-safe feature extractors based on their ability to approximate different C-correlators. Moreover, we concentrate on the extracted features rather than the final observable approximated by the complete network, i.e. we do not consider the function approximation done by the downstream MLP, which takes the extracted IRC safe features, as this would be akin to a usual multi-variate approach of physics motivated features. 

As we will study the general behaviour of the approximated function whose weights are frozen after some training procedure, we will not discuss the explicit dependence of the neural networks on their tuneable parameters in the following discussions.

\subsection{Energy Flow Networks}
Energy Flow Networks are infra-red and collinear safe deep sets model which learns a per-particle map of each particle's directional coordinates $\mathbf{\hat{p}}_i$ and undergoes an energy-weighted sum to form a fixed length representation of any variable cardinality constituent set. 
Without loss of generality for a multi-dimensional representation, a single IRC safe observable can be written as 
\begin{equation*}
	\label{eq:efn}
	C_1 = \sum_{i} z_i\;g_1(\mathbf{\hat{p}}_i) \quad, 
\end{equation*} 
where $g_1(\hat{p}_i)$ represents a parameterised multilayer perceptron. We have specifically denoted the observable as $C_1$ to make it self-evident the per-particle map essentially approximates any general $\mathcal{C}_1^{f_1}$. This is because the MLP $g_1$ is a universal approximator and can approximate any function $f_1$ suiting a particular objective up to a required precision. In a practical implementation, several related IRC safe observables are approximated, which are fed to a downstream network for classification. The direct implementation of EFNs can, therefore, only extract features expressible in terms of $C_1$.

\subsection{Energy-weighted Message Passing Networks}
An energy-weighted message passing operation for any general parametrised function $\mathbf{\bar{g}}^{(\alpha)}$ can be written as
\begin{equation*}
	\mathbf{h}_{i}^{(\alpha+1)}=\sum_{j\in \mathcal{N}[i]} \; \omega^{(\mathcal{N}[i])}_j\; \mathbf{\bar{g}}^{(\alpha+1)}(\mathbf{h}^{(\alpha)}_i,\mathbf{h}^{(\alpha)}_j) \quad,
\end{equation*} where $\mathbf{h}^{(\alpha)}_i$ is the input node features for the $\alpha^{th}$ message passing operation and $$\omega^{(\mathcal{N}[i])}_{j}=\frac{p_T^{j}}{\sum_{k\in\mathcal{N}[i]}\; p^k_T}\quad$$ are the energy weights dependent on the IRC safe neighbourhood set $\mathcal{N}[i]$, with $\omega^{(S)}_{j}=z_j$, for the whole set $\mathcal{S}$. For notational convenience in the following discussions, we will take the sum over the full set of particles in the jet and replace $z_j$ in place of $\omega^{(\mathcal{N}[i])}_{j}$ without loss of generality. Therefore, we have 
\begin{equation}
	\label{eq:empn} 
	\mathbf{h}_{i}^{(\alpha+1)}=\sum_{j} \; z_j\; \mathbf{g}^{(\alpha+1)}(\mathbf{h}^{(\alpha)}_i,\mathbf{h}^{(\alpha)}_j) \quad
\end{equation}  with the function $\mathbf{g}^{(\alpha+1)}$ expressed as a product of a Heaviside step functions $\Theta(\Delta R_{ij}<R_0)$ and the original message function $\mathbf{\bar{g}}^{(\alpha+1)}$ as 
$$\mathbf{g}^{(\alpha+1)}(\mathbf{h}^{(\alpha)}_i,\mathbf{h}^{(\alpha)}_j)=\Theta(\Delta R_{ij}<R_0)\;\mathbf{\bar{g}}^{(\alpha+1)}(\mathbf{h}^{(\alpha)}_i,\mathbf{h}^{(\alpha)}_j)\quad.$$ 
Here, $\Delta R_{ij}$ is the Euclidean distance in the rapidity-azimuth plane between particle $i$ and $j$ while $R_0$ is the graph's radius. The requirement of symmetry in the argument of $f_2(\mathbf{\hat{p}}_i,\mathbf{\hat{p}}_j)$ for $\mathcal{C}^{f_2}_2$ and its absence in eq.~\ref{eq:empn} is not a contradiction as the node features themselves are defined for each particle and hence are not IRC safe observables. In contrast, the IRC safe graph representation will generally be expressible as some linear combination of $\mathcal{C}^{f_N}_N$.

We have $\mathbf{h}^{(0)}_i=\mathbf{\hat{p}}_i$ which gives $\mathbf{\hat{p}}_i=\mathbf{\hat{p}}_j\implies \mathbf{h}^{(\alpha)}_i=\mathbf{h}^{(\alpha)}_j$ for any $\alpha
\geq0$ and any two collinear particles $i$ and $j$. The IRC safe graph representation is obtained as 
\begin{equation*}
	\mathbf{G}^{(L)}=\sum_{i=1}^{n_{part}}\; z_i\; \mathbf{h}^{(L)}_i\quad, 
\end{equation*} 
after $L$ iterations. As we shall see in the following, the complexity of the extracted features via EMPNs will depend on the value of $L$. 

%Although it was shown in ref~\cite{Konar:2021zdg}, that the network satisfies IRC safety for any general $L$, the networks employed were mostly restricted to $L=1$ for a better comparison with Energy Flow Networks. However, within the EMPN formalism itself, there is considerable room for improvement and increasing complexity of the feature extraction which we first highlight in the following sections. To  do this, we simplify the structure by restricting ourselves to complete so that the 

Explicitly for $L=1$, we have $\mathbf{h}_i^{(1)}=\sum_j z_j\;\mathbf{g}^{(1)}(\mathbf{\hat{p}}_i,\mathbf{\hat{p}}_j)$ which gives 
\begin{equation*}
	\mathbf{G}^{(1)}=\sum_{i,j}\; z_i\; z_j \;\mathbf{g}^{(1)}(\mathbf{\hat{p}}_i,\mathbf{\hat{p}}_j)\quad.
\end{equation*} 
If the symmetry is enforced in $\mathbf{g}^{(1)}$, the approximated observable will contain a $\mathcal{C}_2^{f_2}$ term alone. At the same time, a non-symmetric $\mathbf{g}^{(1)}$ would also have a $\mathcal{C}^{f_1}_1$ component.

For $L=2$, we have 
\begin{equation}
	\label{eq:four_body} 
	\begin{split} 
	\mathbf{G}^{(2)}&=\sum_{i,j}\; z_i\; z_j \;\mathbf{g}^{(2)}(\mathbf{h}^{(1)}_i,\mathbf{h}^{(1)}_j)\\
	\implies \mathbf{G}^{(2)}&=\sum_{i,j}\; z_i\; z_j \;\mathbf{g}^{(2)}(\sum_k z_k\,\mathbf{g}^{(1)}(\mathbf{\hat{p}}_i,\mathbf{\hat{p}}_k), \sum_l z_l \,\mathbf{g}^{(1)}(\mathbf{\hat{p}}_j,\mathbf{\hat{p}}_l) )~.
	\end{split} 
\end{equation}
The complicated nature of the arguments makes it difficult to ascertain the exact behaviour of the functional approximation. One expects the universal approximator $\mathbf{g}^{(2)}$ to be expressible as a linear combination of $\mathcal{C}_N^{f_N}$'s up to $N=4$. However, due to the presence of four angular arguments and four energy weights, it hints against the efficient approximation of any $\mathcal{C}_N^{f_N}$ for any $N<4$. 

The situation is even more futile for $L=3$ with eight angular arguments and eight energy-weighted sums. For a particular $L$, we have $2^L$ angular arguments and the same number of energy-weighted sums. Even if one extracts the graph features at each stage $\alpha$, and gets a concatenated graph representation for each $\alpha>0$ up to $\alpha=L$, we have the efficient extraction of $2, 2^2, 2^3,....2^L$ terms the sum in eq.~\ref{eq:c_corr} for any general IRC safe observable $\mathcal{O}$. Although, for jet substructure applications, one does not need to go to very high $N$, we already run into a problem for top-tagging, which has valuable information in the 3-prong structure of the energy deposits.

\section{Hypergraph Energy-weighted Message Passing Networks} 
\label{sec:hepmn} 
As discussed above, although powerful, Graph Neural Networks cannot look into higher-order relational information amongst the nodes efficiently. Therefore, in this section, we develop IRC-safe point cloud architectures capable of efficiently extracting higher-point correlation.

A possible way to extend the capabilities of IRC safe feature extraction to higher-point correlations is to directly implement the form of C-correlators as
\begin{equation*}
	\mathcal{H}^N=\sum_{i_1}\sum_{i_2}...\sum_{i_N} z_{i_1}z_{i_2}... z_{i_N}\;\Theta_N(\hat{p}_{i_1},\hat{p}_{i_2},...,\hat{p}_{i_N})\,\Phi_N(\hat{p}_{i_1},\hat{p}_{i_2},...,\hat{p}_{i_N})\quad,  
\end{equation*} where $\Theta_N$ are step functions for reducing the sums to localised information, and $\Phi_N$ are the neural networks approximating a correlated set (as the output of $\Phi_N$ in general, is a vector) of $f_N$'s for the particular training objective. For IRC safety, both $\Theta_N$ and $\Phi_N$ should be symmetric under the permutation of its arguments. The step function $\Theta_N$ for each $N$ essentially endows an $N$-uniform hypergraph structure onto the constituent set similar to the radius filter $\Theta(\Delta R_{ij}<R_0)$ endowing a graph structure for the case of $N=2$. Therefore, the concatenated hypergraph representations $$\mathbf{X}=\oplus_N\mathcal{H}^N\quad,$$ up to $N_{max}$ would extract IRC safe features to be fed to a downstream MLP for some task. 

We do not follow this approach for the following reasons. It is well-known~\cite{10.7551/mitpress/7496.003.0016,10.1145/1390156.1390294,MAL-006,7780459} that automatic feature extraction works best with deeper networks. Depth can only be brought into $\Phi_N$ in the above expression, which does nothing to the IRC-safe feature extraction process. The complexity can be increased by increasing $N$, which increases the width of the network, thereby increasing the model complexity sharply. Although the factorisation of the extracted features in energy and angular components could lead to better all-order behaviour in QCD and is indeed interesting, one needs to have proper control of the behaviour of the parameter optimisation before we can hope to answer such questions as demonstrated in reference~\cite{Kasieczka:2020nyd}.

Our approach is based on one-particle and two-particle messages to construct a hybrid message-passing neural network that can extract higher point correlations in a recursive approach. Although it is easily generalisable to higher-point information, we restrict ourselves up to 3-point interactions due to the increasing complexity. 

\subsection{IRC safety with heterogenous source and destination embeddings}
The basic observation which makes it possible to build a higher-point IRC safe feature extractor is that the requirement of IRC safety for EMPN is still valid even when the node embedding for the source $\psi_S(\mathbf{\hat{p}}_i)$, and destination $\psi_D(\mathbf{\hat{p}}_i)$ are different as long as they are separately equal in the collinear limit of two particles. If a particle $q$ has two collinear daughters $r$ and $s$, then we have
\begin{equation*}
	\psi_S(\mathbf{\hat{p}}_q)=\psi_S(\mathbf{\hat{p}}_r)=\psi_S(\mathbf{\hat{p}}_s)\quad,\text{ and } \psi_D(\mathbf{\hat{p}}_q)=\psi_D(\mathbf{\hat{p}}_r)=\psi_D(\mathbf{\hat{p}}_s)
\end{equation*} when $\mathbf{\hat{p}}_q=\mathbf{\hat{p}}_r=\mathbf{\hat{p}}_s$, even if $\psi_S(\mathbf{\hat{p}}_i)\neq \psi_D(\mathbf{\hat{p}}_i)$. More importantly, the embeddings $\psi_S$ and $\psi_D$ need not be functions of just a single particle. They can also be the updated node features of the $\alpha$-hop IRC safe neighbourhood after $\alpha$ energy-weighted message passing operations (as given in eq.~\ref{eq:empn}).  
For an IRC safe neighbourhood of $i$, where a particle $q$ splits to two daughters $r$ and $s$, we have $\mathcal{N}[i]\ni q\implies \mathcal{N}'[i]\ni r \;\land \mathcal{N}'[i]\ni s$ when $\mathbf{\hat{p}}_q=\mathbf{\hat{p}}_r=\mathbf{\hat{p}}_s$.

Let us look closer into the statement that we need not have the same embedding in the argument of the message function in an Energy-weighted Message Passing operation even though the statement logically follows from the non-requirement of symmetry of the message function. Since we have heterogeneous source and destination embeddings, we need to fix a uniform direction of messages. We will take the direction of all messages as originating from a neighbourhood node $j\in\mathcal{N}[i]$ to the destination node $i$. Therefore, we have

\begin{equation*}
	\label{eq:empn_assym} 
	\mathbf{H}_{i}^{(\alpha+1,\beta+1)}=\sum_{j} \; z_j\; \mathbf{g}^{(\alpha+1,\beta+1)}(\mathbf{h}^{(\alpha)}_{D,i},\mathbf{h}^{(\beta)}_{S,j}) \quad,
\end{equation*}
where $\mathbf{h}^{(\alpha)}_{D,i}$ and $\mathbf{h}^{(\beta)}_{S,j}$ are the destination and source node embeddings, respectively, and $\mathbf{g}^{(\alpha+1,\beta+1)}$ is the corresponding message function.  
As the destination and source node embeddings differ, the message-passing operations are indexed separately with $\alpha$ and $\beta$, respectively.  
The source embedding satisfying $\mathbf{h}^{(\beta)}_{S,q}=\mathbf{h}^{(\beta)}_{S,r}=\mathbf{h}^{(\beta)}_{S,s}$ in the collinear limit makes the updated node representation $\mathbf{H}_{i}^{(\alpha+1,\beta+1)}$ equal for $i\notin \{q,r,s\}$, in the splitted and unsplitted case since $z_q=z_r+z_s$. Explicitly, we have 
\begin{equation}
	\label{eq:lr_eq}  z_q\;\mathbf{g}^{(\alpha+1,\beta+1)}(\mathbf{h}^{(\alpha)}_{D,i},\mathbf{h}^{(\beta)}_{S,q})=z_r\;\mathbf{g}^{(\alpha+1,\beta+1)}(\mathbf{h}^{(\alpha)}_{D,i},\mathbf{h}^{(\beta)}_{S,r})+z_s\;\mathbf{g}^{(\alpha+1,\beta+1)}(\mathbf{h}^{(\alpha)}_{D,i},\mathbf{h}^{(\beta)}_{S,s})\quad.
	\end{equation}   Additionally, we require the equality of the destination embeddings $\mathbf{h}^{(\alpha)}_{D,q}=\mathbf{h}^{(\alpha)}_{D,r}=\mathbf{h}^{(\alpha)}_{D,s}$ when $i\in \{q,r,s\}$.
 However, we can have $\mathbf{h}^{(\alpha)}_{D,q}\neq \mathbf{h}^{(\beta)}_{S,q}$, as this is not needed to satisfy eq.~\ref{eq:lr_eq}. Therefore, $\mathbf{H}^{(\alpha+1,\beta+1)}_i$ satisfies $\mathbf{H}^{(\alpha+1,\beta+1)}_q=\mathbf{H}^{(\alpha+1,\beta+1)}_r=\mathbf{H}^{(\alpha+1,\beta+1)}_s$, in the collinear limit of the two daughters $r$ and $s$ of $q$. 
 
 \subsection{Building higher point IRC safe feature extractor}
 It is now straightforward to build an IRC-safe message-passing operation which looks into three-particle correlations. The structure of the two-particle energy-weighted operation is kept the same as eq.~\ref{eq:empn}, and then combined with destination embedding $\psi_D(\mathbf{\hat{p}}_i)$ and source embedding $\psi_S(\mathbf{\hat{p}}_i)$  of the angular coordinates to give an effective three particle message passing of the form
 \begin{equation}
 	\begin{split} 
 	\mathbf{H}^{(1,2)}_{i}&=\sum_j\;z_j\; \mathbf{g}^{(1,2)}(\psi_D(\mathbf{\hat{p}}_i),\mathbf{h}^{(1)}_{S,j})\quad, \\
 	\mathbf{H}^{(2,1)}_{i}&=\sum_j\;z_j\; \mathbf{g}^{(2,1)}(\mathbf{h}^{(1)}_{D,i},\psi_S(\mathbf{\hat{p}}_j))  \quad.
 	\end{split} 
 \end{equation}
 As the destination and source embeddings are different, $\mathbf{h}^{(1)}_{D,i}$ and $\mathbf{h}^{(1)}_{S,i}$ denote node features updated after two separate message-passing operations as given in eq.~\ref{eq:empn} with different message functions $\mathbf{g}^{(1)}_D$ and $\mathbf{g}^{(1)}_S$, respectively. The IRC safe feature would be a graph-level representation after an energy-weighted summed graph readout  on $\mathbf{H}^{(1,2)}_i$ and $\mathbf{H}^{(2,1)}_i$, as
 \begin{equation}
 		\mathbf{G}^{(1,2)}_3 = \sum_i \;z_i\; \mathbf{H}^{(1,2)}_i\quad,\quad\mathbf{G}^{(2,1)}_3 = \sum_i \;z_i\; \mathbf{H}^{(2,1)}_i\quad.
 \end{equation}We shall see in the following discussions that these two representations look at distinct topological structures in the graph; the IRC safe representation for the order three feature extraction is constructed as a concatenation of these two components $$\mathbf{G}_3=\mathbf{G}^{(1,2)}_3\oplus\mathbf{G}^{(2,1)}_3\quad.$$
We can ascertain the behaviour of $\mathbf{G}_3$ by writing down its dependence  on the particle's four vectors:
\begin{equation*}
	\begin{split} 
		\mathbf{G}_3 &= \sum_{i,j} \;z_i\,z_j\;\left(\mathbf{g}^{(1,2)}(\psi_D(\mathbf{\hat{p}}_i),\sum_l\;z_l\,\mathbf{g}^{(1)}_S(\mathbf{\hat{p}}_j,\mathbf{\hat{p}}_l))\right.\\
		&\quad\quad\quad\quad\quad\quad\oplus\left.\mathbf{g}^{(2,1)}(\sum_l\,z_l\,\mathbf{g}^{(1)}_D(\mathbf{\hat{p}}_i,\mathbf{\hat{p}}_l),\psi_S(\mathbf{\hat{p}}_j)\right)\quad. 
		\end{split} 
	\end{equation*}
Three energy weights and three angular arguments hint that the learning procedure would directly start looking at the three-particle interrelations. It is important to note that any IRC safe observable looking into $n$ body phase space, by definition, approaches its $n-1$ body phase space limit when one particle approaches the soft or collinear limit. In other words, eq.~\ref{eq:four_body} will also look into the three-body limit of any four-particle combination when one is soft or collinear to any other particle. However, we expect the above form to extract better the three-particle correlations required for tagging three-prong jets like top quarks.

\begin{figure}[t!] 
	\centering
	\includegraphics[scale=0.4]{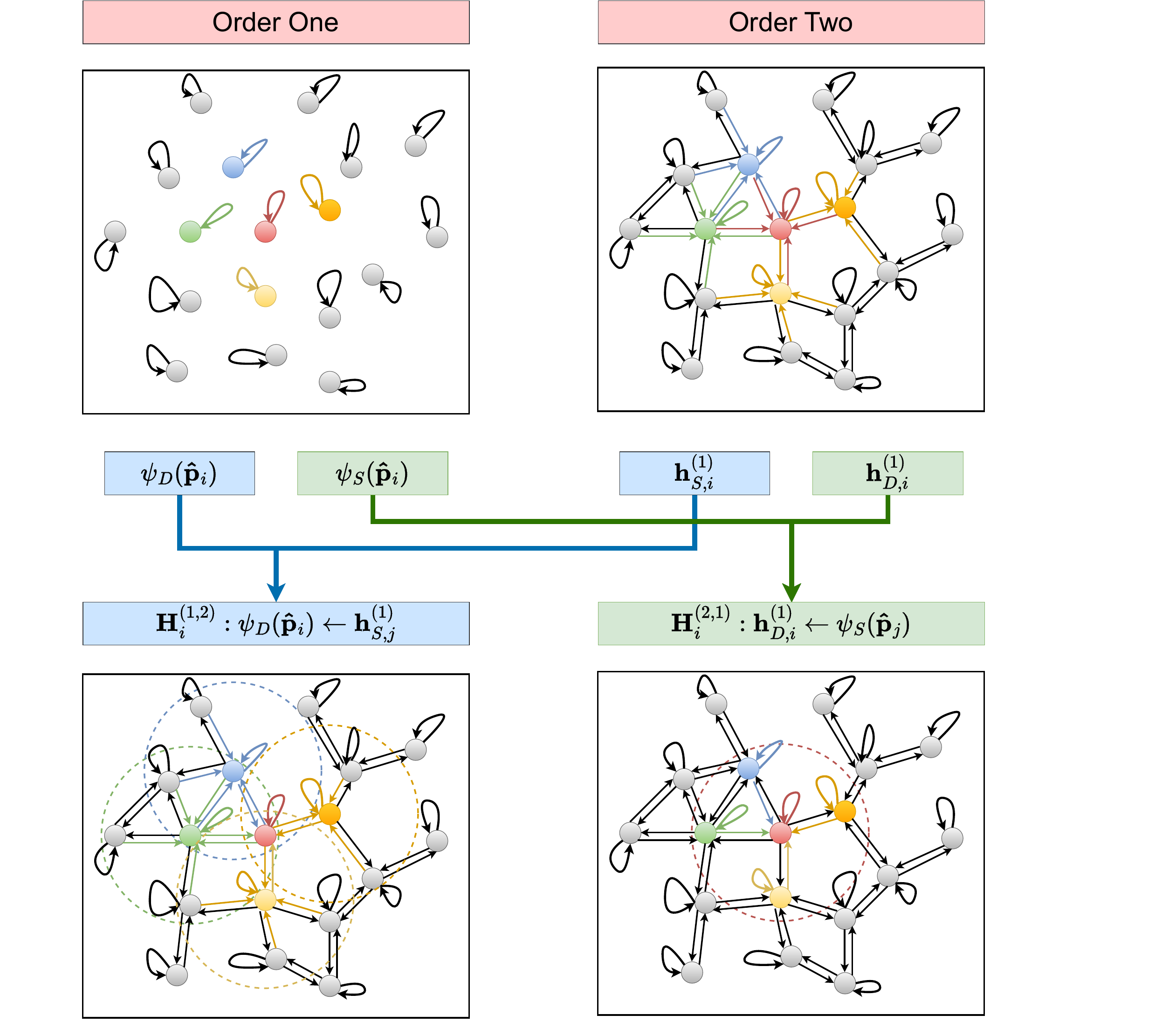}
	\caption{The figure shows a schematic representation of the message passing operation to build hybrid order three node representations for Hypergraph Energy-weighted Message Passing Networks by combining order one and two node representations.}
	\label{fig:hempn} 
\end{figure}

A schematic representation of the feature extraction procedure using different source and destination embeddings of order one and order two operations is shown in figure~\ref{fig:hempn}. We focus on the red node whose neighbours are the coloured. On the top left, the per-particle embeddings for the source and destination can only look into the individual particle information. On the right, however, the energy-weighted message-passing operation gathers information from each node's neighbourhood, which are shown with the identically coloured arrows for the coloured nodes. The order three feature extractors are built by combining the per-particle destination embedding with the order-two source embedding (on the left) and the order-two destination embedding with the per-particle source embedding (on the right).

From a feature extraction perspective, there are two essential differences in comparison to the $L=2$ case given in eq.~\ref{eq:four_body}:
\begin{itemize}
	\item One argument in both $\mathbf{g}^{(1,2)}$ and $\mathbf{g}^{(2,1)}$ is an embedding of the angular coordinates of a single particle and hence contain single-particle information. In contrast, both arguments already contain the aggregated neighbourhood information in  $\mathbf{g}^{(2)}$. 
	\item The embedding of the two arguments in $\mathbf{g}^{(1,2)}$ and $\mathbf{g}^{(2,1)}$ have independently trainable weights while they are shared for $\mathbf{g}^{(2)}$. 
\end{itemize} 
The first difference makes it possible for the function $\mathbf{g}^{(1,2)}$ to effectively extract the relation of node $i$ with the updated neighbourhood information of its neighbours (2-hop neighbourhood of $i$), while the function $\mathbf{g}^{(2,1)}$ looks at the aggregated node feature of $i$'s immediate neighbourhood with individual nodes in the same neighbourhood. The difference is also seen in figure~\ref{fig:hempn}, where on the left $\mathbf{H}^{(1,2)}_i$ looks into the features of the nodes within each coloured circle with the red node, while on the right, $\mathbf{H}^{(2,1)}_i$ looks into the feature of the aggregated neighbourhood information of the red node with the individual nodes within its neighbourhood. This essential difference in the feature extraction procedure makes it imperative to devise the two separate message-passing operations as they need to extract topologically different features within the graph.

It is straightforward to generalize this procedure to any arbitrary $N$, with substantial flexibility to choose the extractor guided by the requirement to divide $N$ into two parts in any possible way. Any feature extractor looking into less than $N$ correlations can be used to extract features from topologically distinct paths of length $N$ within the graph. Due to the different combinatorial factors involved, the complexity rises relatively fast with increasing $N$, and we restrict our discussion to $N=3$.

To look into the learnt features of the order one and two feature extractors, we define the graph representation as a concatenation of the source and destination embeddings as 
\begin{equation}
	\begin{split} 
	\mathbf{G}_1 &=\mathbf{G}_{D,1}\oplus\mathbf{G}_{S,1}=\sum_i \;z_i\; (\psi_D(\mathbf{\hat{p}}_i)\oplus\psi_S(\mathbf{\hat{p}}_i))  \quad,\\ 
	\mathbf{G}_2 &=\mathbf{G}_{D,2}\oplus\mathbf{G}_{S,2}=\sum_i \;z_i\; (\mathbf{h}^{(1)}_{D,i}\oplus\mathbf{h}^{(1)}_{S,i})  \quad.
	\end{split}  
\end{equation} 
This gives the concatenated graph readout to be fed to the classifier network as 
\begin{equation}
	\mathbf{G}=\mathbf{G}_1\oplus\mathbf{G}_2\oplus\mathbf{G}_3\quad. 
\end{equation}

 \section{Network architecture and training} 
 \label{sec:arch} 
  \begin{figure}[t!] 
 	\centering
 	\includegraphics[scale=0.3]{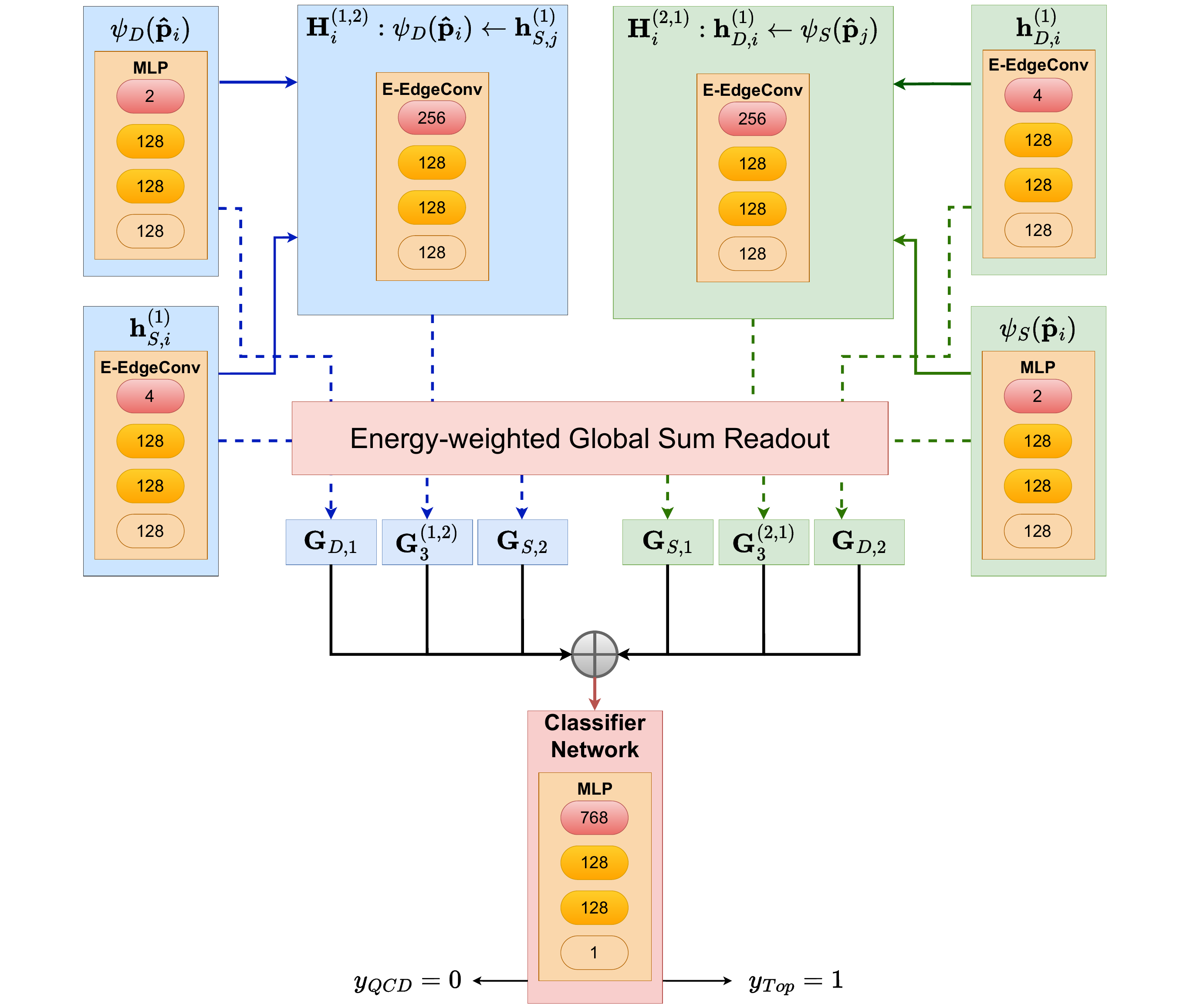}
 	\caption{The architecture of the H-EMPN network utilized in this study is shown as a flowchart.}
 	\label{fig:b_hempn} 
 \end{figure}
 
 To gauge the properties of the proposed network, we utilise the public top-tagging dataset~\cite{kasieczka_gregor_2019_2603256} for a supervised classifier. These events were generated with {\tt Pythia 8.2.15}~\cite{Sjostrand:2014zea} and were showered and hadronised without MPI effects. The showered events additionally underwent a parametrised detector response via {\tt Delphes3}~\cite{deFavereau:2013fsa} with the default ATLAS detector card. The particle-flow objects of the Delphes output were used as inputs to construct anti-$k_T$~\cite{Cacciari:2008gp} jets with $R=0.8$ via {\tt FastJet}~\cite{Cacciari:2011ma}. with additional requirements of $p_T$ within the range $[550,650]$ GeV, and pseudorapidity $|\eta|<2$. Further, for the signal events, the top quark and its decay products' parton level information were used to reject falsely reconstructed jets with the partons falling outside the jet's area. The training data comprises 1.2 million samples, while the test and validation datasets contain 400k samples. The network analysis uses \textsc{PyTorch-Geometric }~\cite{Fey/Lenssen/2019}.

We compare order three Hypergraph Energy-weighted Message Passing Networks (H-EMPNs) with $L=2$ EMPNs. For a reasonable comparison with the H-EMPN, we will extract the graph features for $\alpha=1$ and $\alpha=2$ stages separately for the EMPN and feed the concatenated graph representation into the classifier network. As shown in figure~\ref{fig:b_hempn}, the IRC-safe feature extractor module for the H-EMPN, in total, contains two per-particle maps for $\psi_D$ and $\psi_S$, and four energy-weighted edge convolution  (E-EdgeConv) operations to give the updated node embeddings $\mathbf{h}^{(1)}_{D,i}$, $\mathbf{h}^{(1)}_{S,i}$, $\mathbf{H}^{(1,2)}_{i}$, and $\mathbf{H}^{(2,1)}_{i}$. Including the classifier MLP, which takes in the concatenated graph readout, we have seven MLPs. We have one for each per-particle map and a message function for each E-EdgeConv operation from the feature extractor module. All these seven MLPs contain two hidden layers with 128 nodes and a rectified linear unit activation function. Except for the classifier network, which has a one-dimensional output with sigmoid activation, all other MLPs have a 128-dimensional output layer with a linear activation function. The per-particle maps take the rapidity-azimuth coordinates $\mathbf{\hat{p}}_i=(\Delta y_{iJ},\Delta \phi_{iJ})$ of each constituent $i$ as inputs with the differences taken from the jet axis defined by the four-vector $p^\mu_J=\sum_{k=1}^{n_{part}} p^k_\mu$. For a destination node embedding $\mathbf{h}_{S,i}$ and source node embedding $\mathbf{h}_{D,i}$, the message function takes in the concatenated vector $\mathbf{h}_{S,i}\oplus\mathbf{h}_{S,i}-\mathbf{h}_{S,j}$ as the input. The EMPN network sequentially applies the E-EdgeConv operation twice to the input graph's node features. The first and the second E-EdgeConv operations have the same MLP architecture corresponding to the ones that give $\mathbf{h}^{(1)}_{D,i}$ (or $\mathbf{h}^{(1)}_{S,i}$) and $\mathbf{H}^{(1,2)}_{i}$ (or $\mathbf{H}^{(2,1)}_{i}$), respectively. The classifier MLP for the EMPN and H-EMPN takes in 256  and 768-dimensional concatenated graph representations, respectively. The whole network is trained using the binary-cross entropy loss function.   
   
 We construct graphs with $R_0\in\{0.4,0.5,0.6\}$ and $R_0\to\infty$ corresponding to complete graphs.\footnote{ Strictly speaking, the maximum value that $R_0$ can take is determined by the jet's diameter as we are always confined to particles contained in the jet.  We use the $R_0\to\infty$ limit for defining the complete graph, as we are using sequential recombination algorithms and the maximum area of the jet is not compactly defined even for the anti-$k_t$ algorithm which gives almost conical jets.} For all these four instances of input graphs, we train each network five times from random initialization for 100 epochs with the \texttt{Adam} optimizer~\cite{DBLP:journals/corr/KingmaB14} and a learning rate of 0.001. A decay-on-plateau condition is applied to the learning rate with a decay factor of 0.5 if the validation loss does not decrease for three epochs. The epoch with minimum validation loss is used for inference for each training instance.  
 
\section{Results} 
\label{sec:results} 

\begin{figure}[t!]
	\centering
	\includegraphics[scale=0.18]{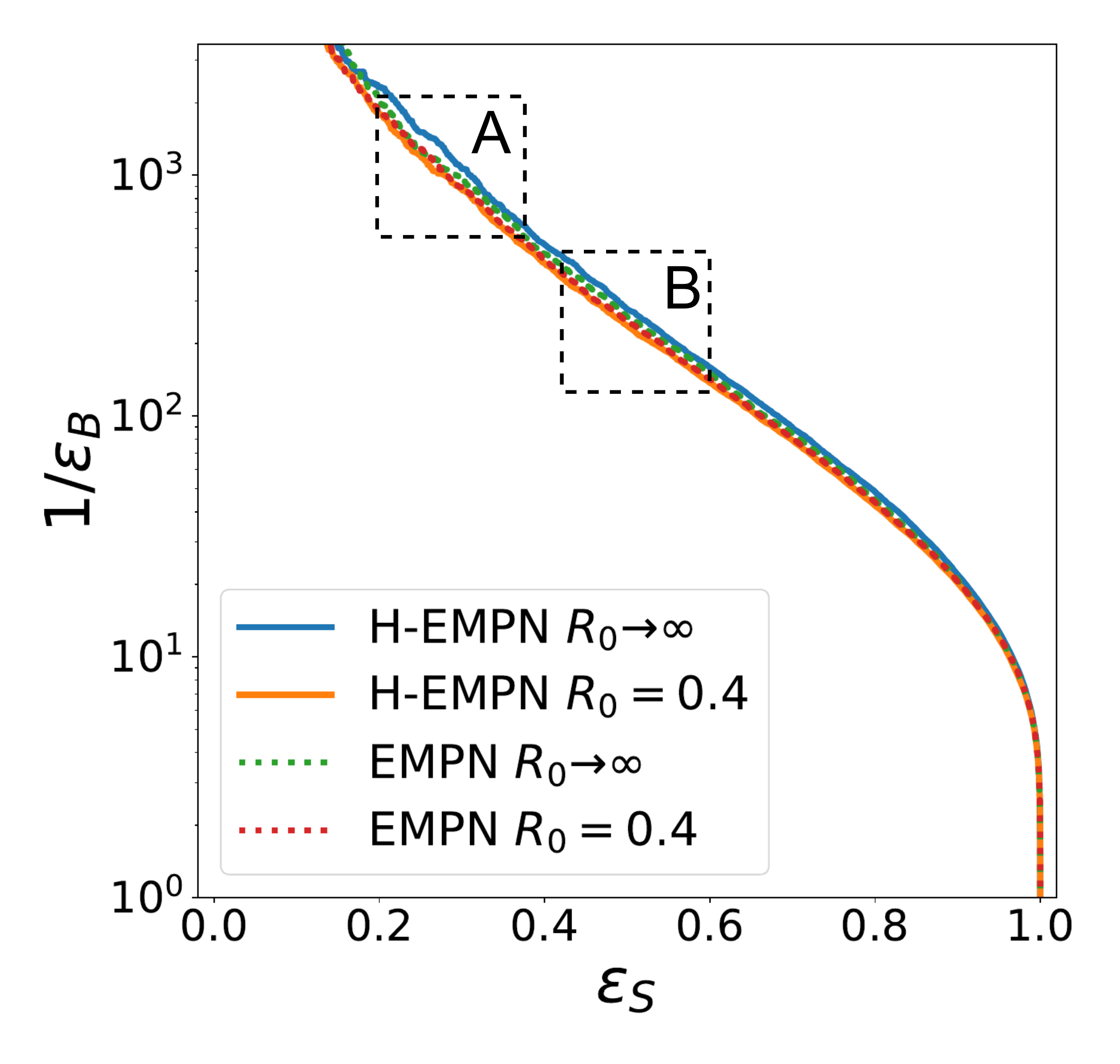}\\
	\includegraphics[scale=0.12]{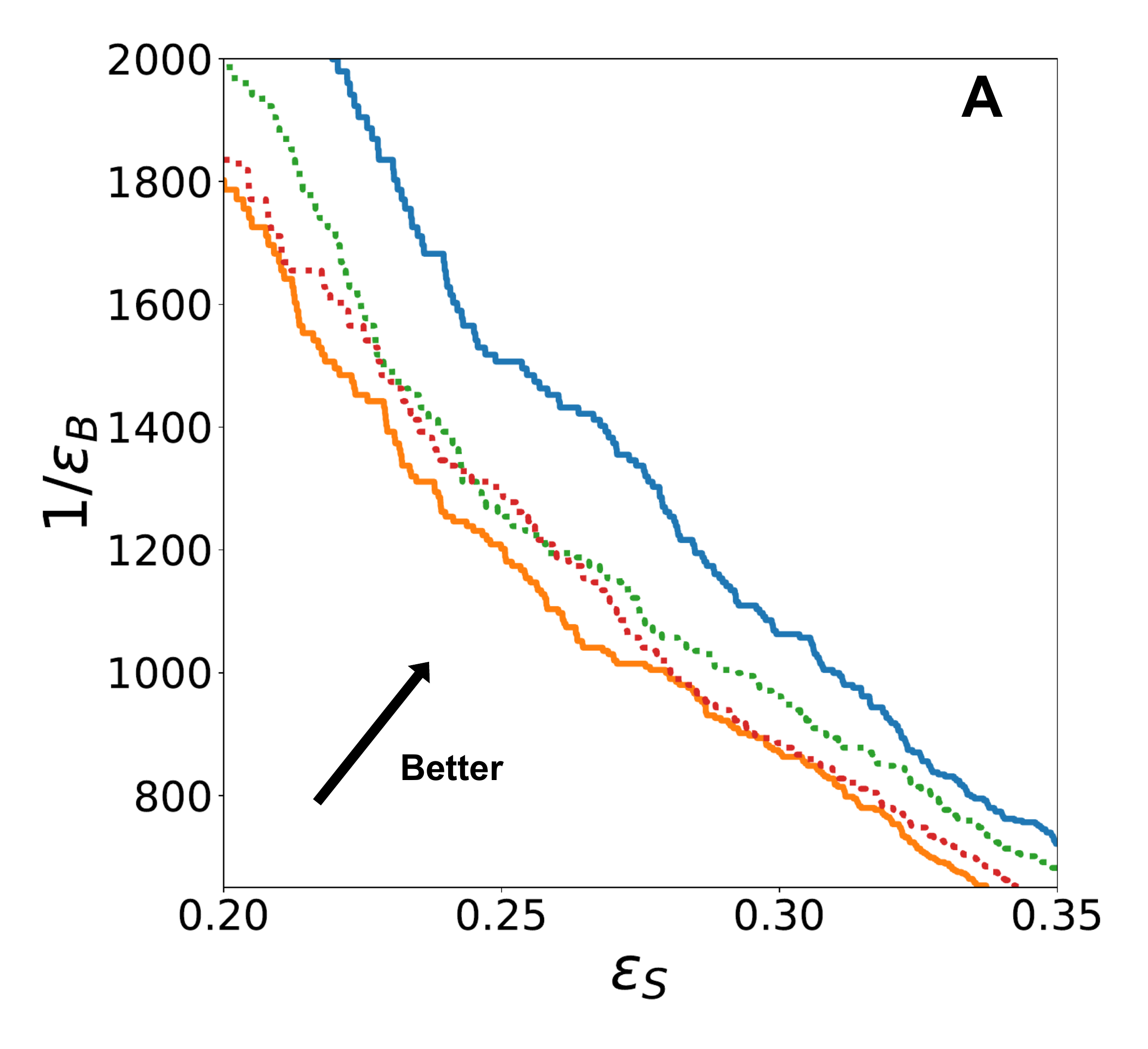}
	\includegraphics[scale=0.12]{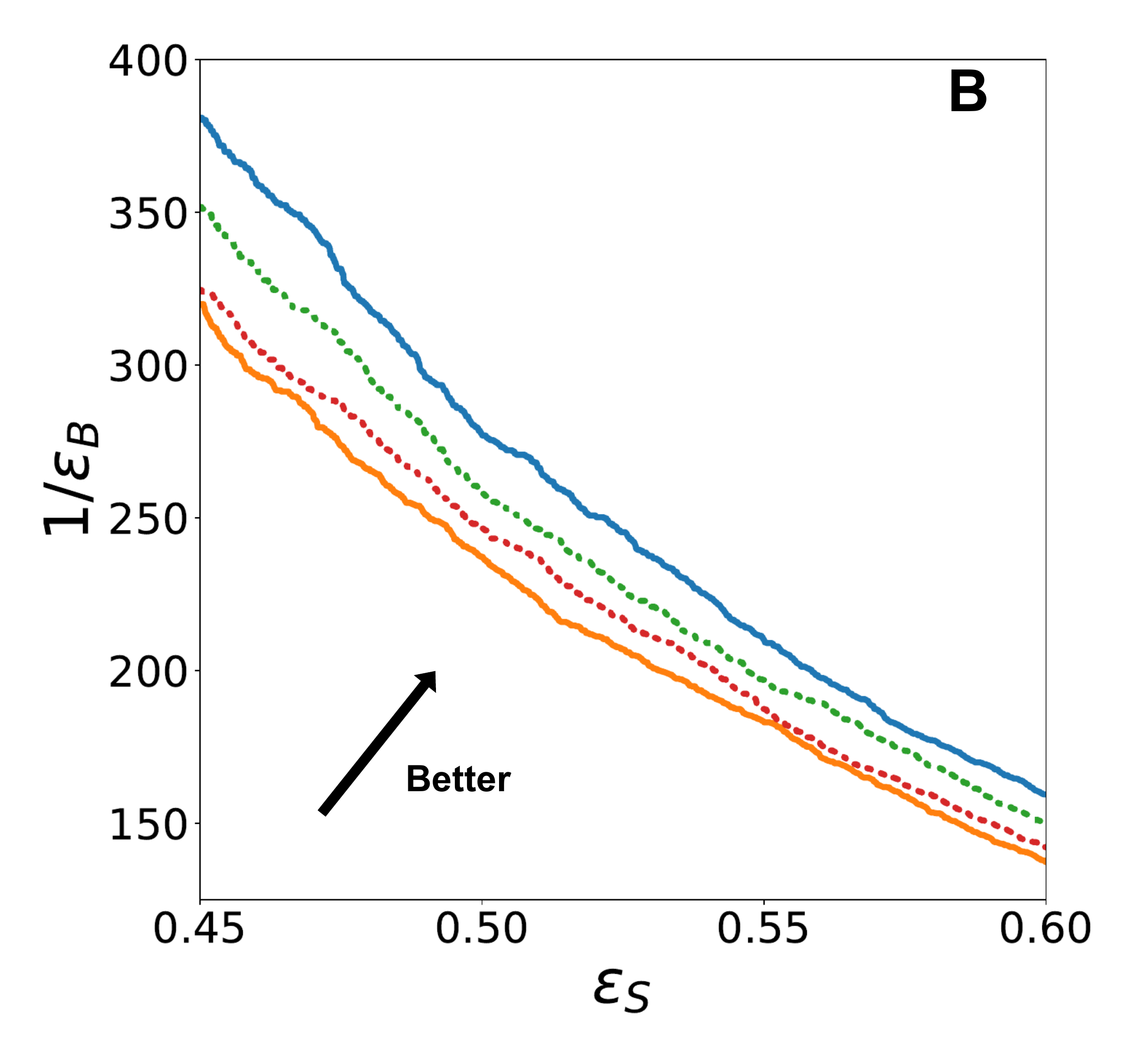} 
	\caption{The receiver operator characteristics curve for the best performing network (in terms of AUC) over the five training instances for $R_0=0.4$ and $R_0\to\infty$ for the EMPN and H-EMPN for different ranges of signal acceptance $\epsilon_S$.  On the left, we show $1/\epsilon_b$ in log scale over the full range of $\epsilon_S$, while on the center and right, it is shown in linear scale over different regions of $\epsilon_S$ to highlight the differences. }
	\label{fig:roc} 
\end{figure}
\subsection{Performance}
The receiver operator characteristics (ROC) curve for the network with highest area under the ROC (AUC) curve from all training instances between the signal acceptance $\epsilon_S$ and the inverse of background acceptance $1/\epsilon_B$ for the two models for $R_0=0.4$ and $R_0\to\infty$ is shown in figure~\ref{fig:roc}. We see that the EMPN has almost an overlapping ROC curve for these two radii, while for the H-EMPN, there is a noticeable improvement. 
The area under the receiver operator curve for the EMPN and H-EMPN for different graph construction radii are tabulated in table~\ref{tab:aucs}. The values correspond to the mean over the five training instances, while the errors correspond to the standard deviation. For $R_0=0.4$, the EMPN and H-EMPN have almost identical discrimination power with an AUC of 0.9823 and 0.9821, respectively.  As the radius increases, there is a steady increase for the H-EMPN, while for the EMPN, it increases for $R_0=0.5$ and stays at a similar value for $R_0=0.6$ and there is a noticeable dip in performance when going to complete graphs with $R_0\to\infty$. This trend can be understood from the structural difference between the EMPN and H-EMPN and the three-prong nature of the top jet. The EMPN's feature extraction is sequential, with the second E-EdgeConv being fed by the first E-EdgeConv's updated node features. With increasing radius, the feature-extraction, which looks at aggregated two-particle correlations, suffers from a redundancy of the information as the first E-EdgeConv already looks at a much larger neighbourhood in the rapidity-azimuth plane. On the other hand, the H-EMPN has a much larger width, with four modules taking the input jet constituents parallelly, which are then combined non-trivially to feed the order-three feature extractors.   Even though the order-three extractors take in the updated order-two node features from the full jet in the $R_0\to\infty$ limit, the combination with the per-particle maps drives the extraction process to look at any relevant three-prong structure in the whole jet.  From a purely QCD perspective, the radius $R_0$ puts in an additional scale beyond the jet radius, and going to the $R_0\to\infty$  limit takes away this dependence in the feature extraction procedure. 
	Although it is possible to define $R_0$ as a function of the IRC safe kinematic information of the jet which could possibly improve the feature extraction, we do not consider this as our aim is to move towards theoretically transparent ways of improving feature extraction. Therefore, the H-EMPN can extract features from the full jet more efficiently without being restrained by an arbitrary angular scale $R_0$.

The AUC paints a global picture of the discrimination power of a binary classifier; however, a classifier is almost always used at a specific working point, depending on the analysis. This practical aspect demands a local figure of merit, which we show with the inverse of the background acceptance $\epsilon_B$, the background rejection $1/\epsilon_B$, at fixed values of signal acceptance $\epsilon_S$. The background rejection for the EMPN and H-EMPN for the different graph construction radii are shown for $\epsilon_S=0.5$ and $\epsilon_S=0.3$ in tables \ref{tab:R50} and \ref{tab:R30}, respectively. The values are averaged over the five training instances, with the standard deviations shown as errors. Although the trend for separate models is similar to that of the AUCs, the H-EMPN already starts having a noticeably better background rejection for $R_0=0.5$ even though the EMPN has a nominally higher AUC. As a matter of fact, except for $R_0=0.4$ at $\epsilon_S=0.3$, the H-EMPN has a numerically higher mean background rejection for all other instances.
\begin{table}[t!]
	\centering 
	\resizebox{1.0\textwidth}{!}{
		\renewcommand{\arraystretch}{1.4}
		\begin{tabular}{|l|c|c|c|c|}
			\hline 
			&\multicolumn{4}{c|}{\textbf{Area Under the ROC Curve}}\\
			\cline{2-5} 
			\textbf{Model}&$R_0=0.4$&$R_0=0.5$&$R_0=0.6$&$R_0\to\infty$\\
			\hline 
			EMPN &$0.9823\pm0.00015$&$0.9827\pm0.00009$&$0.9826\pm0.00024$&$0.9825\pm0.00015$\\
			H-EMPN&$0.9821\pm0.00012$&$0.9826\pm0.00010$&$0.9828\pm0.00029$&$0.9834\pm0.00012$\\
			\hline 
		\end{tabular}
	}\caption{The table shows the mean AUC for five training instances evaluated on the test dataset of the public top-tagging dataset for different architectures. The errors shown are the standard deviation of the five training instances.	}\label{tab:aucs}
\end{table}
\begin{table}[t!]
	\centering 
	\resizebox{0.6\textwidth}{!}{
		\renewcommand{\arraystretch}{1.4}
		\begin{tabular}{|l|c|c|c|c|}
			\hline 
			&\multicolumn{4}{c|}{$1/\epsilon_B$ at $\epsilon_S=0.5$}\\
			\cline{2-5} 
			\textbf{Model}&$R_0=0.4$&$R_0=0.5$&$R_0=0.6$&$R_0\to\infty$\\
			\hline 
			EMPN &$235\pm7$&$250\pm2$&$246\pm4$&$255\pm6$\\
			H-EMPN&$236\pm2$&$258\pm6$&$258\pm11$&$276\pm6$\\
			\hline 
		\end{tabular}
	}\caption{The table shows the background rejection at a signal acceptance of 50\% for different models. The values correspond to the mean from the evaluation of the test dataset for five different training instances from random initialization, while the standard deviations are shown as errors. 	}\label{tab:R50}
\end{table}
\begin{table}[t!]
	\centering 
	\resizebox{0.6\textwidth}{!}{
		\renewcommand{\arraystretch}{1.4}
		\begin{tabular}{|l|c|c|c|c|}
			\hline 
			&\multicolumn{4}{c|}{$1/\epsilon_B$ at $\epsilon_S=0.3$}\\
			\cline{2-5} 
			\textbf{Model}&$R_0=0.4$&$R_0=0.5$&$R_0=0.6$&$R_0\to\infty$\\
			\hline 
			EMPN &$819\pm39$&$882\pm11$&$839\pm31$&$895\pm36$\\
			H-EMPN&$817\pm33$&$917\pm25$&$911\pm34$&$995\pm48$\\
			\hline 
		\end{tabular}
	}\caption{The table shows the background rejection ($1/\epsilon_B)$ at a signal acceptance ($\epsilon_S$) of 30\% for different models. The values correspond to the mean from the evaluation of the test dataset for five different training instances from random initialization, while the standard deviations are shown as errors. }\label{tab:R30}
\end{table}
\subsection{Visualizing the latent graph representation}
    
\begin{figure}[t!]
	\centering
	\includegraphics[scale=0.25]{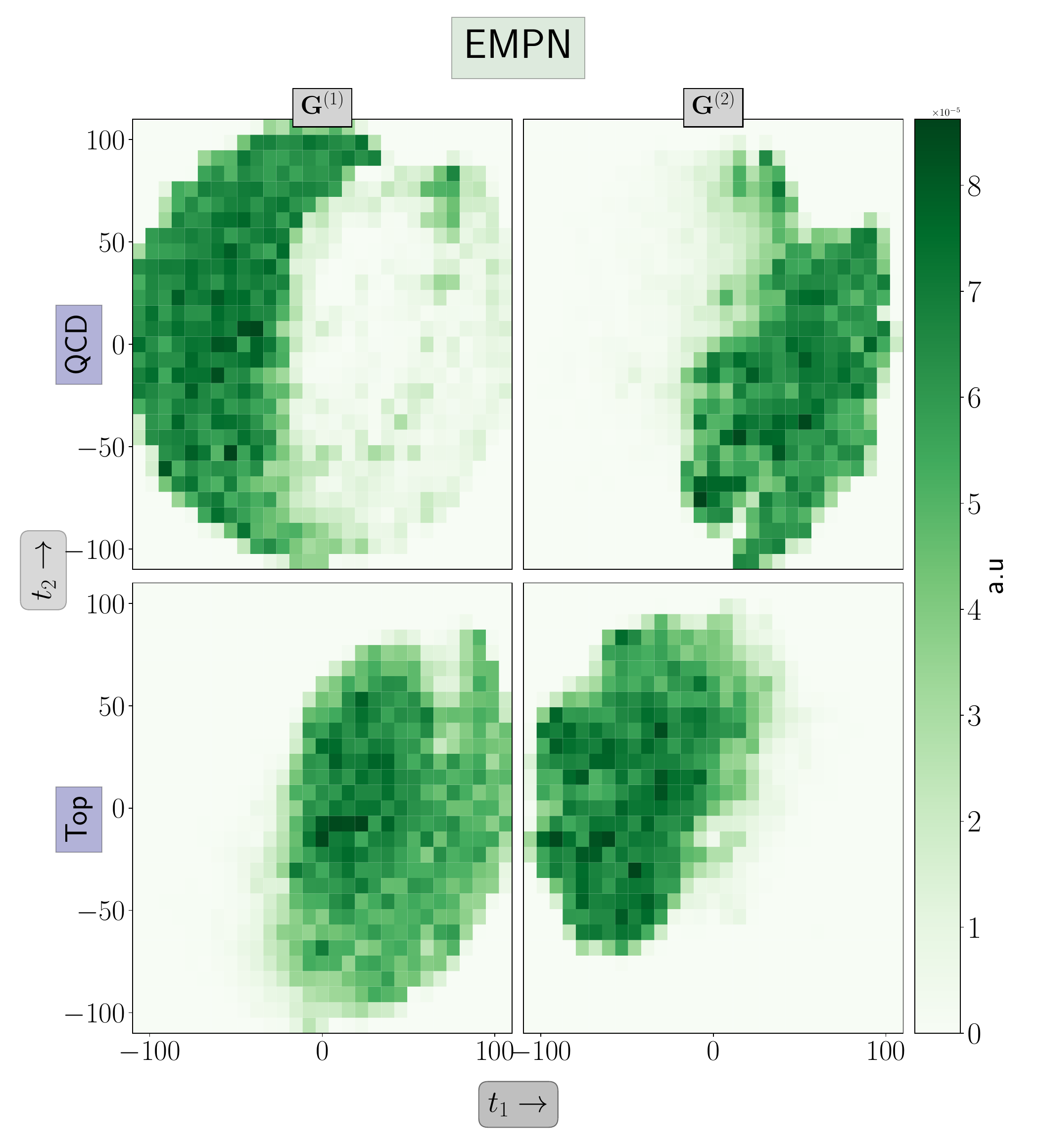}
	\caption{The two-dimensional histogram of the QCD (above) and top (below) test datasets in the two-dimensional latent space obtained after a t-SNE embedding of the 128-dimensional graph representation $\mathbf{G}^{(1)}$ (left) and $\mathbf{G}^{(2)}$ (right) of the best performing EMPN trained with complete graphs.}
	\label{fig:tsne_empn} 
\end{figure} 
\begin{figure}[t!]
	\centering
	\includegraphics[scale=0.25]{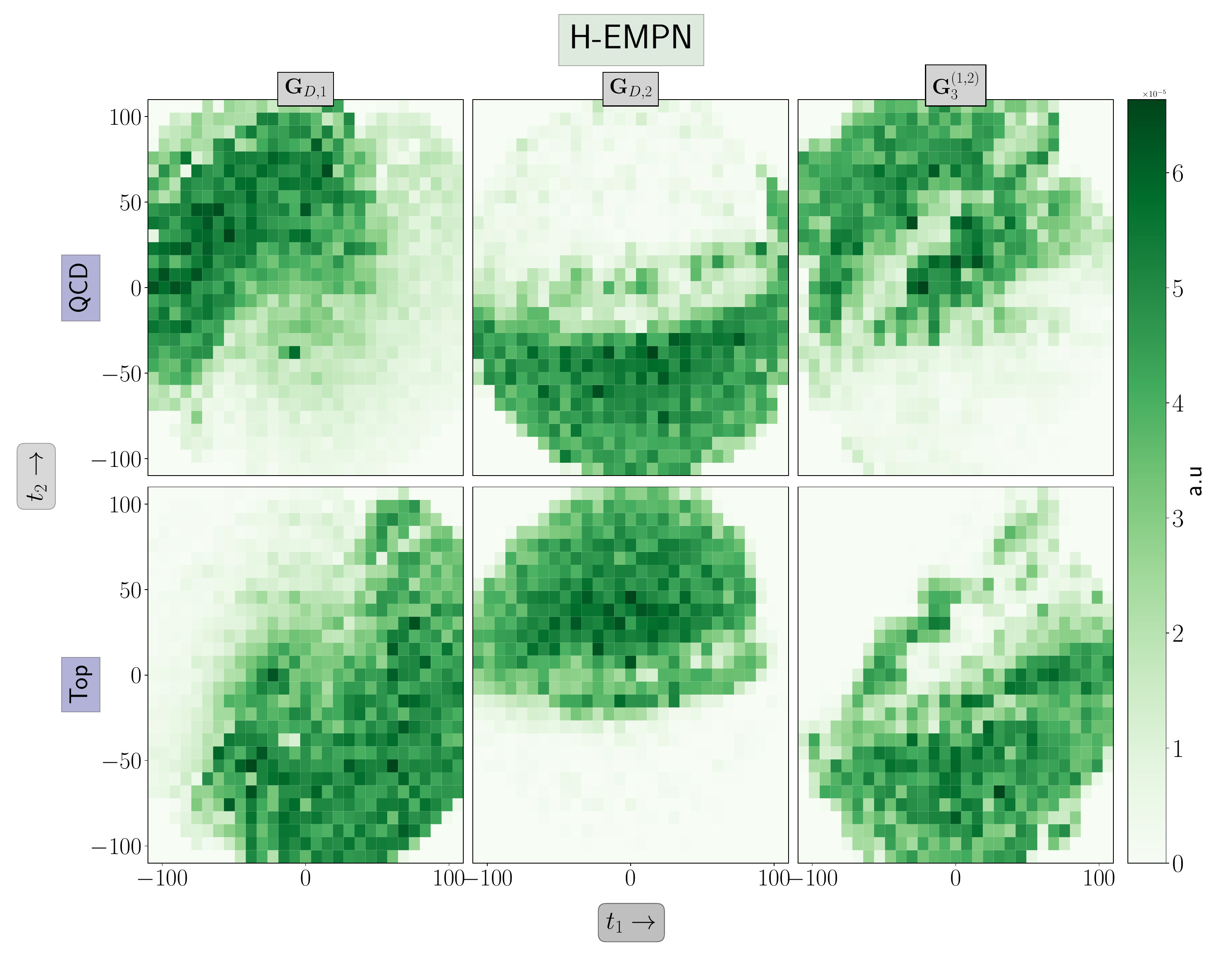}
	\caption{The two-dimensional histogram of the QCD (above) and top (below) test datasets in the two-dimensional latent space obtained after a t-SNE embedding of the 128-dimensional graph representation $\mathbf{G}_{D,1}$ (left), $\mathbf{G}_{D,2}$ (center) and $\mathbf{G}^{(1,2)}_3$ (right) of the best performing H-EMPN trained with complete graphs.}
	\label{fig:tsne_hempn_d} 
\end{figure} 

\begin{figure}[h]
	\centering
	\includegraphics[scale=0.25]{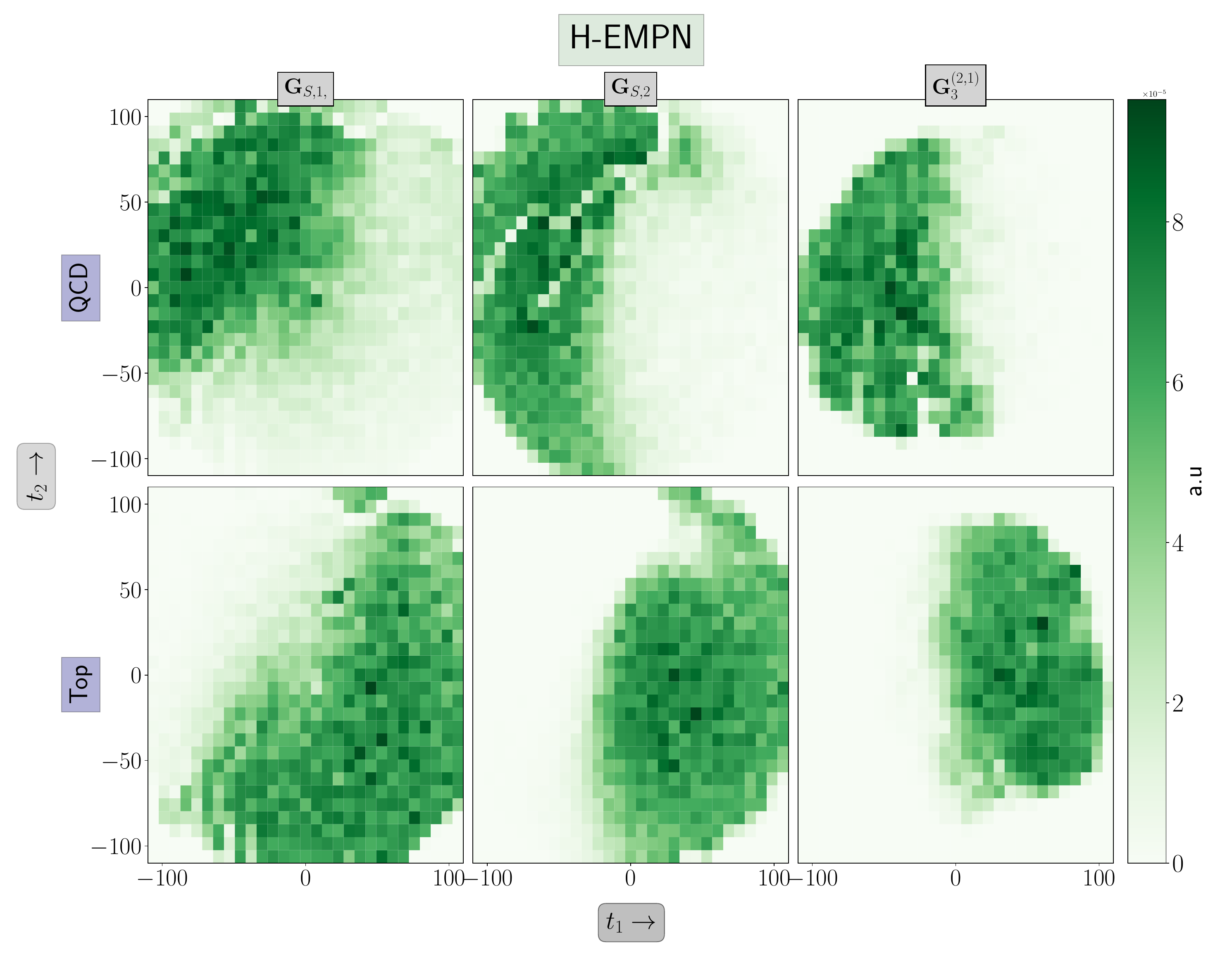}
	\caption{The two-dimensional histogram of the QCD (above) and top (below) test datasets in the two-dimensional latent space obtained after a t-SNE embedding of the 128-dimensional graph representation $\mathbf{G}_{S,1}$ (left), $\mathbf{G}_{S,2}$ (center) and $\mathbf{G}^{(2,1)}_3$ (right) of the best performing H-EMPN trained with complete graphs.}
	\label{fig:tsne_hempn_s} 
\end{figure}

In this section, we investigate whether all the graph representations that the H-EMPN learns can contribute to separating the signal and the background for the final classifier output. We choose the best-performing complete graph, which has the possibility of the highest information redundancy besides being the strongest classifier. Although a relatively high linear correlation with the network output does point to the classification using that particular information, it is defined for each component of the graph representation, which dilutes the importance of the underlying vector representations. Moreover, the absence of linear correlation does not imply the lack of discriminatory information, as neural networks can be highly non-linear functions of their inputs. 

We look into the separating power of the different graph representations by visualizing them in a two-dimensional latent space using the t-distributed Stochastic Neighbourhood Embedding (t-SNE)~\cite{JMLR:v9:vandermaaten08a}---an unsupervised data representation technique, where high dimensional data is embedded non-linearly in a lower dimensional space by maximally conserving the neighbourhood information endowed by a Euclidean metric in both spaces. In other words, nearby points in the high-dimensional representation get mapped to a local neighbourhood in the low-dimensional space. As it is an unsupervised technique, no explicit class information (QCD and top for our case) is fed when learning the map, and the clusters that arise in the low-dimensional space are a consequence of their proximity in the high-dimensional space. Therefore, a well-separated cluster in the lower-dimensional space implies that the higher-dimensional space also has well-separated regions.

We use the implementation of t-SNE in \texttt{Scikit-learn}~\cite{scikit-learn} package to embed the various 128-dimensional graph representations of the test dataset evaluated on the best performing EMPN and H-EMPN for the complete graph in a two-dimensional space separately for each representation. The class-wise two-dimensional histogram in the embedding space $(t_1,t_2)$ for $\mathbf{G}^{(1)}$ and $\mathbf{G}^{(2)}$ for the EMPN are shown in figure~\ref{fig:tsne_empn}. We can see that both the graph representations have relatively distinct regions in $(t_1,t_2)$ for the QCD samples (shown above) and top samples (shown below). Similarly, the two-dimensional histograms for the graph representations constructed out of the destination and source node-embeddings for the H-EMPN are shown in figures~\ref{fig:tsne_hempn_d} and \ref{fig:tsne_hempn_s}, respectively. All these embedded graph representations exhibit clear clustering of the QCD and top samples in different regions, confirming that the H-EMPN has extracted discriminating features from all of its component modules. 

Although the EMPN and H-EMPN can utilize their constituent graph representation to separate the QCD jets from top jets as seen from these two-dimensional histograms, we reiterate the qualitative differences between these two networks from the QCD perspective. The $L=2$ EMPN looks up to order four relations. In contrast, the H-EMPN in its present guise only looks up to order three--the sequential application of E-EdgeConv (to give $\mathbf{H}^{(1,2)}_i$ and $\mathbf{H}^{(2,1)}_i$) takes in the per-particle map with single particle information rather than an updated node feature with the local neighbourhood information in one of its arguments. However, we can see the better ability of the H-EMPN network from its performance studies and potentially better behaviour in QCD with its greater efficacy in the absence of an arbitrary angular scale $R_0$. Since we took the top vs QCD jets classification example, we already knew that there is beneficial information in the three-prong structure within the jet, which prompted our design of the specific  H-EMPN.\footnote{The situation may be different, for instance, in the quark vs gluon case where the separating information is not in the hard prong structure but the soft radiation pattern surrounding the one prong core within the jet.}  The first observation from the finite $R_0$ cases is that the H-EMPN architecture is more critical in extracting the order three relational information from the jets than the $L=2$ EMPN. On the other hand, our a priori knowledge of QCD, prompting the design of the H-EMPN, validates that physical inductive biases, or more specifically, QCD, have an important role in the design of performant feature extractors. Therefore, rather than throwing a currently ``fashionable network" under the hood, designing architectures based on the underlying physical intuition can help push the performance boundaries of deep learning algorithms and gain (at least) a qualitative understanding of their inner workings.

\section{Conclusions} 
\label{sec:conc} 

This study delved deep into the intricacies of generalised automatic infrared and collinear safe feature extraction for LHC phenomenology, focusing on the potential of Graphs and Hypergraphs. Hypergraphs are a generalisation of traditional graphs. While a standard graph consists of vertices connected by edges, each connecting exactly two vertices, a hypergraph allows edges to connect any number of vertices, offering a more flexible way to represent relationships between entities. 

First, we explored the behaviour of energy-weighted message passing and its capability to approximate general infrared and collinear safe observables. We highlighted the significance of IRC-safe observables, especially in the context of data interpretation at  LHC experiments.
The study further explored the capabilities of Energy Flow Networks and Energy-weighted message-passing networks, shedding light on their potential and constraints utilising the usage of multilayer perceptrons as universal function approximators within the architecture with the IRC-safe observables expressible in terms of C-correlators.

To enhance the capabilities of IRC safe feature extraction, especially for higher-point correlations, a novel method was introduced by leveraging the form of C-correlators and heterogenous source and destination node embeddings. This approach presents a renewed outlook on feature extraction.

Qualitatively assessing the two models, while the EMPN model provides a robust foundation for feature extraction, the H-EMPN model, designed to look at order-three interparticle relations, demonstrates an edge in performance metrics even though the EMPN model via the application of two-message passing operations could theoretically look up to order-four. This suggests that incorporating hypergraph structures in the H-EMPN model offers enhanced capabilities in extracting higher-point correlations, making it a promising tool for more intricate analyses in LHC phenomenology.

Our findings underscore the potential of hypergraph-based methods in enhancing the extraction of IRC-safe features. The research paves the way for further exploration into LHC phenomenology, focusing on optimising feature extraction techniques. 

\section*{Acknowledgements}
M.S. is supported by the STFC under grant ST/P001246/1. Computational work were performed on the Param Vikram-1000 High Performance Computing Cluster and TDP resources at the Physical Research Laboratory (PRL).

\bibliographystyle{JHEP}
\bibliography{ref}

\end{document}